STRUCTURE AND DYNAMICS OF THE FERROELECTRIC RELAXORS
$Pb(Mg_{1/3}Nb_{2/3})O_3$ AND $Pb(Zn_{1/3}Nb_{2/3})O_3$

Gen Shirane                                        Peter M. Gehring
Department of Physics                  NIST Center for Neutron Research
Brookhaven National Laboratory    National Institute of Standards and
Upton, NY  11973-5000                Technology
                                                     Gaithersburg, MD  20899-8562

ABSTRACT
   We review some of the current research on two of the ultra-high piezoelectric relaxors $Pb(Mg_{1/3}Nb_{2/3})O_3$ (PMN) and $Pb(Zn_{1/3}Nb_{2/3})O_3$ (PZN).  The discovery of a monoclinic phase in $Pb(Zr_{1-x}Ti_x)O_3$ (PZT) in 1999 by Noheda *et al.* forced a reassessment of the structural symmetries found near the morphotropic phase boundary in PZT, as well as in the $PbTiO_3$ (PT) doped systems PZN-PT, and PMN-PT.  All three systems are now known to have nearly identical phase diagrams that exhibit a rhombohedral-monoclinic-tetragonal structural sequence with increasing $PbTiO_3$ concentration.  The dynamical properties of these relaxors are controlled by the unique nature of the so-called polar nanoregions (PNR), which first appear at the Burns temperature $T_d$ as evidenced by the onset of diffuse scattering.  That this diffuse scattering persists to temperatures below the Curie temperatures of PMN (220 K) and PZN (410 K) indicates that the phase transitions in these relaxors are very different from those observed in ordinary ferroelectrics.  They also differ in this regard from what would be expected in the often-quoted model based on analogy with the random field effect.

INTRODUCTION
   Current relaxor research spans a wide range of technical and scientific areas, from industrial applications, to basic experimental measurements, to first-principles theoretical calculations. The relaxor systems PZN-PT and PMN-PT, which have shown the most promise in the area of device applications, were discovered during the early 1980's [1].  However impressive advances in our understanding of the origin of their exceptional piezoelectric properties have only been achieved during the last several years. These advances are summarized in a series of up-to-date review articles [2-5].  These reviews reference a large number of recent publications. Consequently we have targeted the focus of this review on a few recent developments that we believe are changing the direction of research on relaxors.

With respect to structural studies, the monoclinic phase first discovered in PZT [6-8] has now been documented in the two key relaxor systems PZN-PT and PMN-PT, as shown in Fig.1. Some details of these structural aspects will be in the next section.

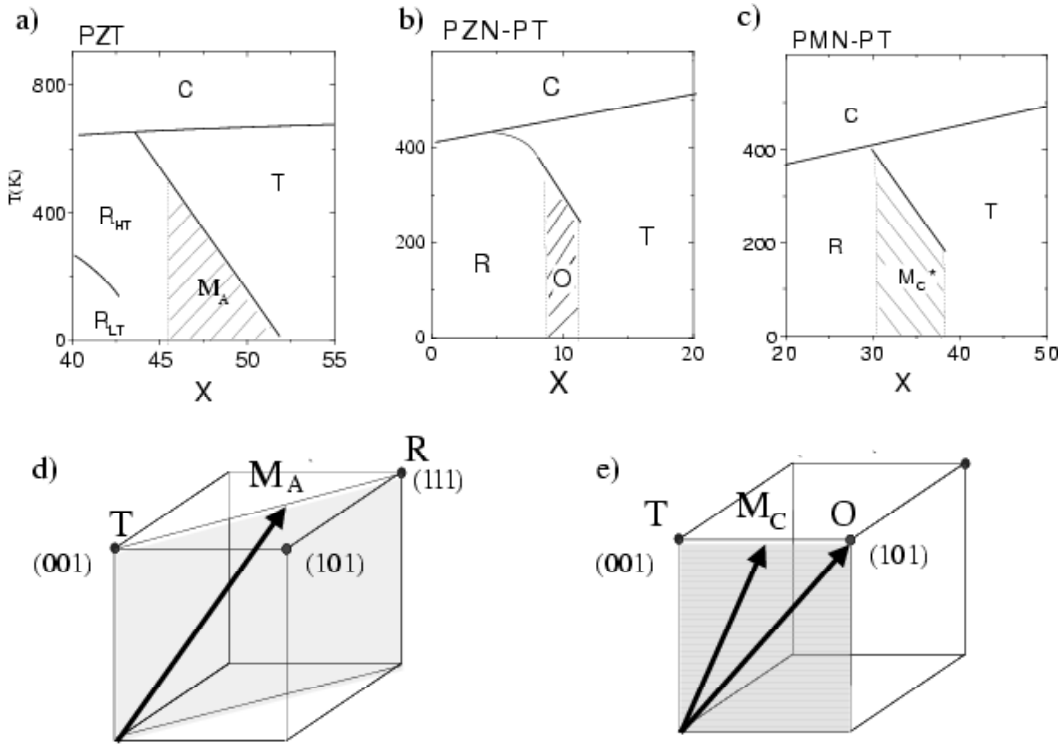

Fig. 1 Phase diagrams are shown for a) PZT, b) PZN-PT, and c) PMN-PT. The hatched regions indicate the $M_A$ and $M_C$ phases located near the MPB. From Noheda [3].

Gehring *et al*. have summarized the dynamical aspects of relaxor phase transitions in a recent review [2]. The existence of a true ferroelectric soft mode in these relaxor systems is now conclusively established after a series of neutron scattering experiments on PZN-PT, PZN, and PMN [9-12]. The centerpiece here is the existence of the so-called polar nanoregions, or PNR. Burns and Dacol first proposed the existence of these unique entities, which are intrinsic to relaxors, in a seminal paper [13]. The onset of the diffuse scattering at the Burns temperature $T_d$, and thus its connection to the PNR, was first reported in the pioneering paper by Naberezhnov *et al*. [14]. Hirota *et al*. later related the static atomic arrangements within the PNR to the ionic vibrations of associated with the soft TO mode through an analysis of the diffuse scattering [15]. A detailed discussion of this is given in a later section. It appears now that all of the important characteristics of relaxors are attributable to the persistence of the PNR both above and below the Curie temperature $T_c$.

MONOCLINIC PHASES NEAR THE MPB

The existence of a narrow region of monoclinic phase located between the rhombohedral and tetragonal phases was first discovered in ceramic samples of PZT [8]. Later detailed structural studies were carried out for the PZN-PT system, where Park and Shrout [16] reported ultrahigh piezoelectric behavior. As shown in panel (b) of Fig.1, the monoclinic phase was subsequently discovered in this system as well [17-19]. This phase is actually orthorhombic, but the $M_C$ phase is extremely close in energy, and a very small electric field E//[001] is sufficient to induce the monoclinic phase where the polarization can rotate between the O and T phases shown in panel (e) [3]. Very recently, a similar phase diagram was reported for PMN-PT [20-21].

Vanderbilt and Cohen [22] highlighted the key features of these phase diagrams in lucid fashion using a phenomenological theory. They demonstrated that only tetragonal, orthorhombic, and rhombohedral structural distortions appear when the free energy expansion is limited to $6^{th}$ order. Indeed, only when one includes the $8^{th}$ order term do the monoclinic phases $M_A$ and $M_C$, shown in the lower part of Fig. 1, appear in the phase diagram. Both of these new monoclinic phases have already been established as one can see in the top figures. The orthorhombic phase in PZN-PT can be regarded as the end member of the $M_C$ phase. Actually one crystal of $M_C$ symmetry was found in PZN-9%PT [18]. The $M_A$ phase in PZT was also derived by Bellaiche *et al*. [23] using the first-principles theoretical approach.

As emphasized by these theoretical papers, these new monoclinic phases appear only in extremely anharmonic crystals; thus they rarely exist in nature. Because of the nearly flat free-energy surfaces, these monoclinic phases can easily be transformed into other nearby phases either by electric field or mechanical stress. This is probably closely related to the great potential of these materials in industrial applications.

This phase degeneracy is manifested in many different ways in experimental observations. One example of this degeneracy is shown in the x-ray diffraction data of Fig. 2 obtained from a powder sample of PMN-35%PT [24]:

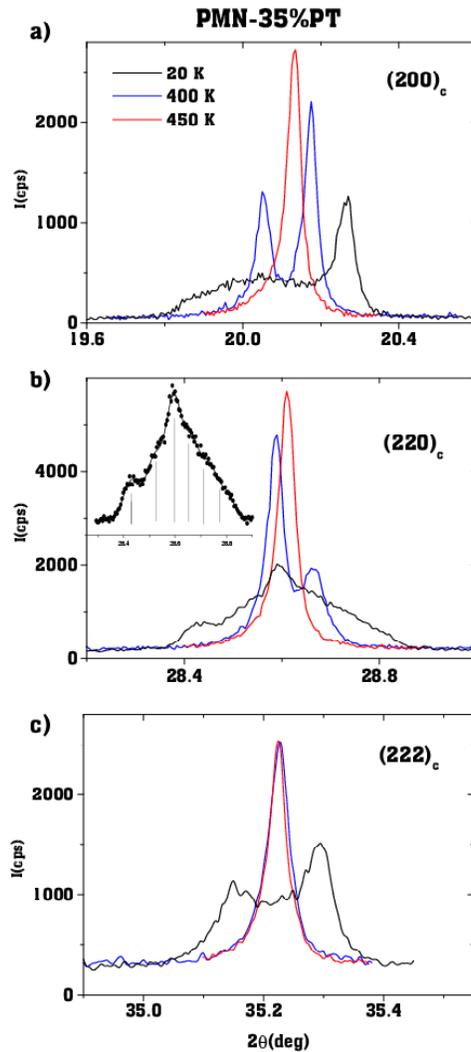

Fig. 2 X-ray profiles of powdered PMN-35%PT. From Noheda *et al*. [24].

This particular composition goes through a cubic-tetragonal-monoclinic phase sequence on cooling. As one can see clearly in Fig. 2, the x-ray profiles in the cubic (450 K) and tetragonal (400 K) phases are very sharp. Only in the monoclinic phase do the profiles become very broad. This is not due to a co-existence of two phases in the usual sense. Rather it is the flatness of the free-energy surface that permits two monoclinic structures, with slightly different lattice parameters, to coexist. In a similar fashion, an optical study of a crystal with a similar $PbTiO_3$ concentration reported finding different polarization directions in different parts of the crystal [25].

STRUCTURAL CHANGES UNDER ELECTRIC FIELD

The structural modifications induced by an applied electric field are, of course, one of the most intensively studied topics on high piezoelectric oxides. One of the more interesting x-ray experiments in this area was reported by Guo *et al*. [26] on ceramic PZT x=42%, which is the rhombohedral phase. The standard concept of a poled ceramic at some elevated temperature involves the reorientation of preferred domains under an electric field as shown in Fig.3:

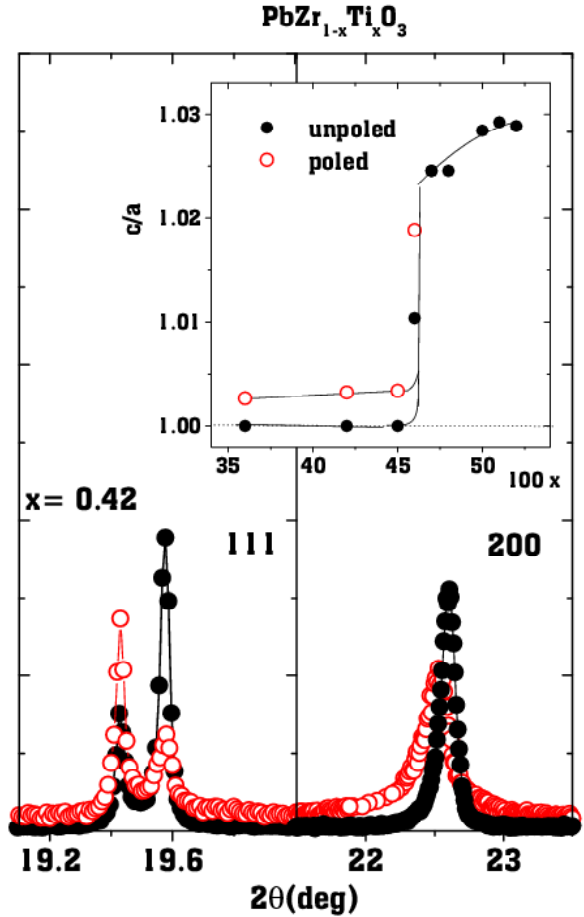

Fig. 3. Poling-induced changes in the x-ray profiles of PZT containing 42% Ti. The positions of the (200) peaks are shifted by poling, and are an indication of the induced, and retained, monoclinic phases. The inset shows unpublished data of Noheda and Guo [27]: this induced monoclinic phase is observed over a wide range of Ti concentration. From Guo *et al*. [26].

The flood of research activity on PZN-PT was sparked by the report of Park and Shrout [16] of giant strains reaching 1.6% by poling the rombohedral concentration of PZN-8%PT along [001]. An extensive series of x-ray studies have been carried out on this crystal [17,19,21,24,28]. In contrast to the PZT system, for which only ceramic samples are available, high-quality single crystals of the relaxor PZN-PT can be grown. Noheda *et al*. [21,28] demonstrated the existence of a skin effect in this system in which the near-surface structure, of order a few microns deep, is strongly modified by the unusually strong piezoelectric coupling. In other words, the true structure of the crystal bulk can be revealed only by the use of sufficiently high-energy x-rays (~60 keV), or by use of neutrons, which by their nature are more highly penetrating than x-rays.

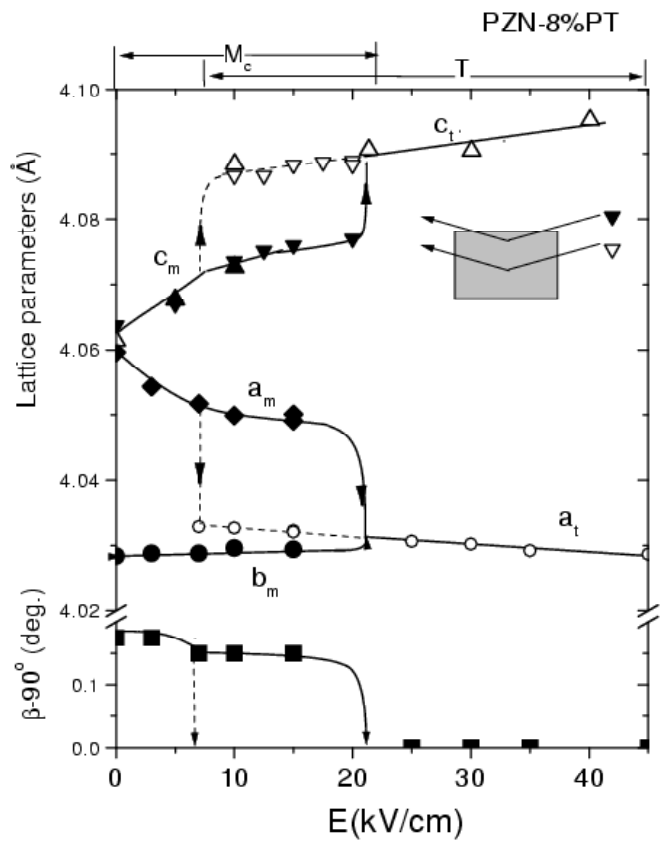

Fig. 4  Structural changes in PZN-8%PT as a function of electric field applied along [001] [24]. The crystal transforms from a high-field tetragonal phase to a low-field monoclinic $M_C$ phase around E = 22 kV/cm. From Noheda *et al*. [24].

As shown in Fig. 4, the polarization sequence for PZN-8%PT involves the monoclinic phase, not the rhombohedral phase. Starting originally from the rhombohedral-to-$M_A$ path, the system jumps over to the $M_C$ path and stays there. This delicate polarization path was studied theoretically by Fu and Cohen [29] and Bellaiche *et al*. [30].

Very recently, Ohwada *et al*. [31] systematically studied the polarization path as a function of temperature and field-cooling history, and mapped out the phase diagram shown in Fig. 5. Their astonishing finding is the creation of the $M_C$ phase with an electric field as low as 500 V/cm applied in the cubic phase. They found that the expected rhombohedral phase is never realized in the field (E // [001]) vs. temperature phase diagram. They discovered a new phase X, which is nearly cubic, that is only established under zero-field cooling. Further discussion regarding this phase X will be given at the end of this review.

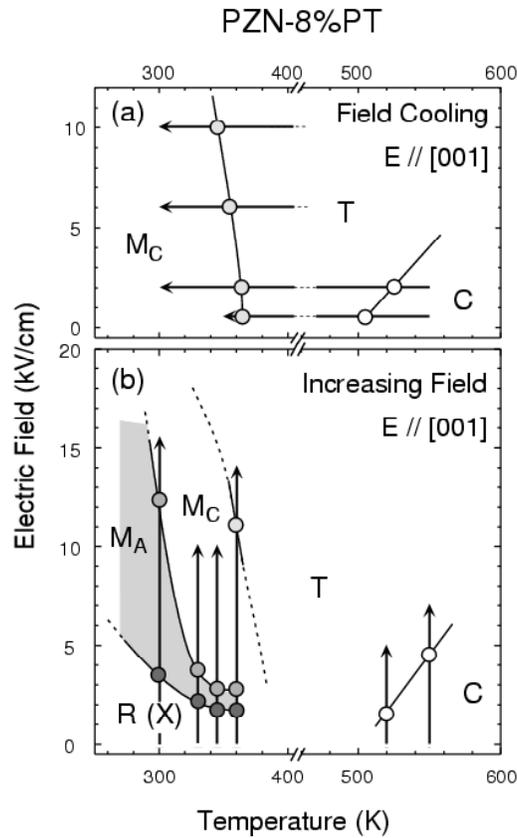

Fig. 5 Field versus temperature phase diagram of PZN-8%PT with E // [001]. The $M_A$-$M_C$ phase boundary was explicitly identified in this study. Once induced, the $M_C$ phase remains after removal of the field. From Ohwada *et al*. [31].

PHONON DISPERSION AND WATERFALL

Different groups have measured the lattice dynamics of PMN at high temperatures using neutron scattering techniques [11,14]. Above the Burns temperature $T_d = 620$ K the dynamics of PMN are similar to those observed in cubic $PbTiO_3$. A well-defined, low-frequency, transverse optic (TO) phonon is observed throughout the Brillouin zone in PMN at 1100 K, which follows the dispersion curve shown in Fig. 6 [11]. The reduced momentum transfer (wave vector) q is measured in reciprocal lattice units (1 r.l.u. = 1.55 Å$^{-1}$) relative to the zone center along the [010] symmetry direction.

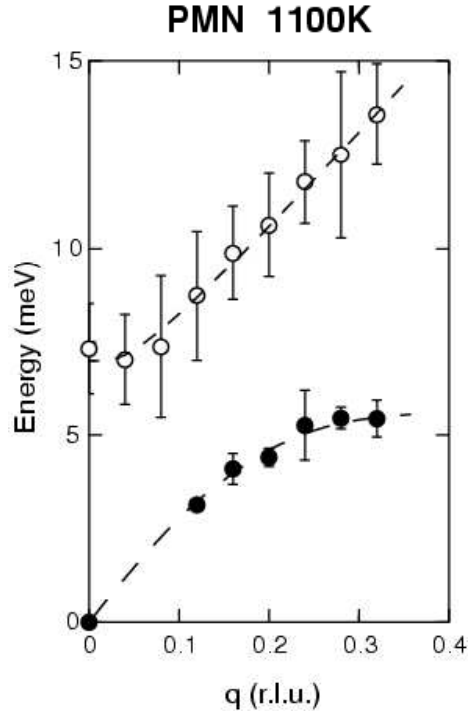

Fig. 6. Transverse optic (TO) (open circles) and transverse acoustic (TA) (solid circles) phonon dispersions in PMN measured at 1100 K along the [010] direction. Vertical bars indicate the phonon linewidths. From Gehring *et al*. [11].

This situation changes with the appearance of the polar nanoregions below $T_d$. The polar nanostructure intrinsic to the relaxors PMN and PZN that forms below the Burns temperature effectively impedes the propagation of long-wavelength TO phonons, and gives rise to a strong wave vector dependent damping. As the phonon lifetimes decrease in response to the growing number of PNR, the corresponding phonon linewidths (which vary inversely with lifetime) increase. This is the explanation for the now well-known "waterfall" feature observed below $T_d$ in PZN, shown in Fig. 7, as well as in PMN, PZN-PT and PMN-PT [9,11].

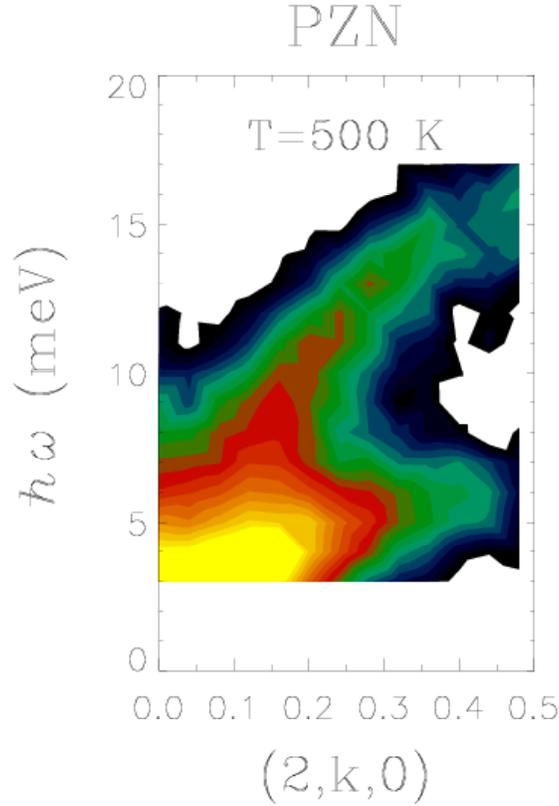

Fig. 7. Logarithmic color contour plot of the neutron inelastic scattering intensity measured in PZN at 500 K in the (200) Brillouin zone. Yellow represents the highest intensity. The vertical red region around $k = 0.14$ r.l.u. corresponds to the waterfall anomaly in which the TO phonon branch *appears* to plummet into the TA branch. From Gehring *et al.* [10].

Below $T_d$ the long-wavelength (low-q) TO phonons become overdamped over a range of q starting from the zone center (q=0). In other words, no phonon peak appears in the neutron scattering spectra at non-zero frequency over this range of q. Instead, the spectral weight of the total scattering function $S(q,\omega)$ is pushed towards the elastic ($\omega=0$) channel, and a simultaneous increase in diffuse scattering is observed [14-15]. For wave vectors larger than a temperature-dependent critical wave vector $q_{wf}$, i.e. for sufficiently small wavelengths, a normal propagating TO phonon mode is observed. Thus $q_{wf}$ is clearly related to the size and density of the PNR. In Fig. 7, $q_{wf}$ is of order 0.14 r.l.u., or 0.20 Å$^{-1}$.

However, the intrinsic value of $q_{wf}$ is difficult to ascertain from measurements of the waterfall because the dynamical structure factor varies from one Brillouin zone to another. For this reason the vertical waterfall anomaly appears centered at different q-values for measurements made in the (200) zone compared to those made in the (300) zone. In the absence of a theoretical description of the phonon scattering cross section that provides an explicit dependence of the phonon linewidth $\Gamma$ on the PNR, it is impossible to make more than a rough estimate of the intrinsic $q_{wf}$, and thus the intrinsic size/density of the PNR. A more quantitative, albeit phenomenological, model is provided by a mode-coupling description that is discussed in a later section.

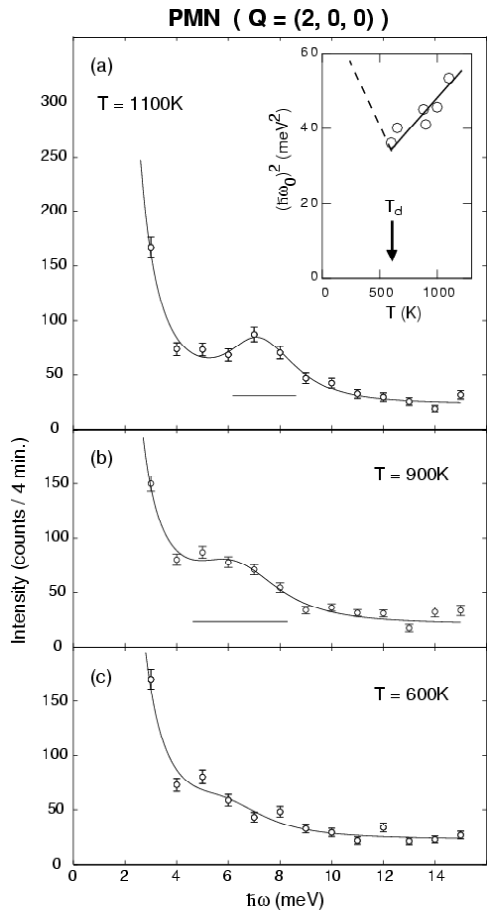

Fig. 8. Zone center TO phonon peak profiles measured at **Q** = (200) at a) 1100 K, b) 900 K, and c) 600 K ($< T_d$). The horizontal bars indicate the intrinsic phonon linewidths. From Gehring *et al.* [11].

DISPLACED POLAR NANOREGIONS

The concept of the polar nanoregion is central to an understanding of the lattice dynamics of the relaxors PMN and PZN. The key questions concern the origin and structure of the PNR. A clue to the origin of the PNR was provided by the neutron scattering measurements by Gehring *et al.* on PMN between 1100 K and $T_d \sim 620$ K, which yielded the first unambiguous evidence of a soft TO mode in this system [11]. The data in Fig. 8 document the evolution of the zone center TO mode upon cooling from 1100 K to just below the Burns temperature. In addition to softening to lower energy with decreasing temperature, one sees that the soft mode broadens rapidly in energy (i.e. the phonon lifetime decreases). Intriguingly, the soft mode becomes so broad near $T_d$ that it is effectively overdamped. This is interesting because the overdamping of the soft mode would then seem to coincide with the formation of the PNR, thereby suggesting that the two phenomena are related. In other words, the PNR may originate from the condensation of the soft mode.

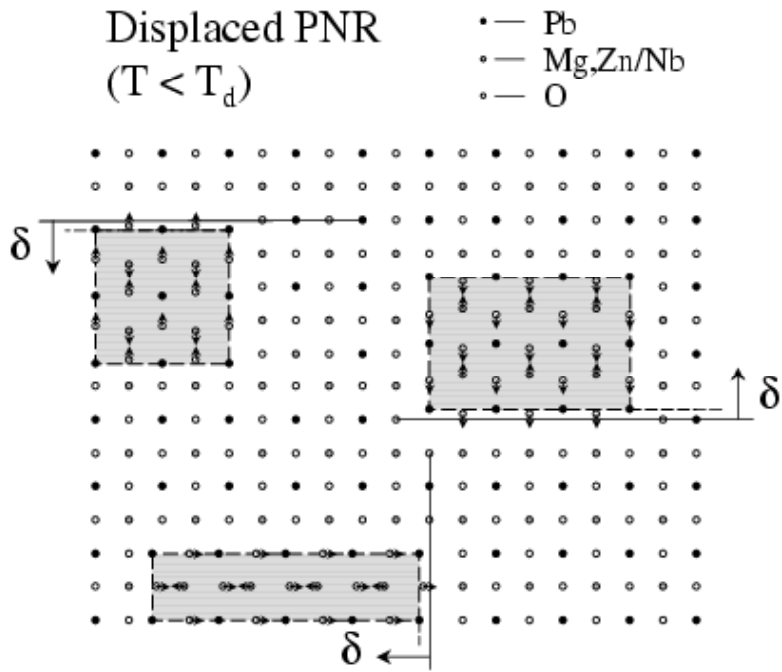

Fig. 9. A schematic representation, suggested by the model of Hirota *et al.*, of how the PNR (shaded regions) are displaced relative to the underlying cubic lattice. Arrows indicate (for simplicity) a Slater-type of tetragonal distortion.

To test this idea, Hirota *et al.* performed a careful analysis of the diffuse scattering in PMN [15]. As mentioned in the introduction, the onset of diffuse scattering in PMN was found to coincide with $T_d$, and thus connected to the PNR [14]. The room temperature ionic displacements in PMN had already been determined in 1995 by Vakhrushev *et al.* from careful neutron diffuse scattering measurements made near 16 different reciprocal lattice positions [32]. Surprisingly, the ionic displacements did not preserve the unit cell center of mass, as is required of any distortion that originates from the condensation of a transverse optic vibration. However Hirota *et al.* noticed that by adding a constant $\delta$ to the displacements of each atom in the unit cell, the remaining shifts could be made to satisfy the necessary center of mass condition. This idea gave birth to the concept of a "phase-shifted condensed soft mode," or equivalently "displaced polar nanoregions," which are depicted schematically in Fig. 9.

One possible method to test the concept of the displaced PNR experimentally would be to measure the diffuse scattering in PMN or PZN above and below $T_c$ both with and without an electric field applied along the [001] direction. If the diffuse scattering measured along the field direction should decrease, while that along the perpendicular direction stays constant, then such an observation would lend further credence to this novel picture. Preliminary data from Ohwada, as yet unpublished, suggest that this may be the case [33].

SOFT MODES BELOW $T_c$

Neutron inelastic experiments by Wakimoto *et al.* on a larger single crystal of PMN significantly extend the previous soft mode measurements to much lower temperatures [12]. The major results from this study are summarized in Fig. 10. A striking finding is that, some 400 K below the Burns temperature, the soft TO mode reappears, i.e. it is no longer overdamped. Further, the square of the soft mode energy varies linearly with temperature, which is the hallmark of a ferroelectric soft mode. The solid black lines shown in the bottom panel of Fig. 10 indicate the range of temperatures over which this ferroelectric behavior was observed. These results represent a dynamical signature of a ferroelectric phase transition, thereby implying that the crystal symmetry is no longer cubic. As such they challenge the accepted view that PMN remains cubic at low temperatures. In this regard it is particularly interesting to note that the soft mode reappears very close to the temperature $T_c = 213$ K because this is the temperature at which the electric-field-induced ferroelectric state in PMN is lost upon zero-field heating [34]. It is tempting to speculate that the low-temperature phase of PMN is the same phase X as that found in PZN by Ohwada *et al.* [31].

The top panel of Fig. 10 shows the temperature dependence of the TA phonon linewidth $\Gamma$. Naberezhnov *et al.* were the first to observe an increase in the TA linewidth near the Burns temperature [14]. The results of Wakimoto *et al*. reveal the remarkable finding that Γ actually peaks, and then decreases below 400 K, before finally recovering essentially the same value at $T_c$ as observed at $T_d$. While these data were taken at a specific value of q = 0.20 r.l.u. in the (200) Brillouin zone, similar results were obtained for wave vectors q = 0.12 and 0.16 r.l.u. Thus the overdamping of the soft mode, and its subsequent recovery, appear to coincide with the anomalous broadening, and then narrowing, of the TA phonon, respectively. The broadening of the TA mode at $T_d$ suggests that the PNR create a non-uniform distortion of the lattice that affects the acoustic phonon lifetimes. However the diffuse scattering in PMN continues to increase below 300 K [14]. Therefore the PNR persist to low temperature. It was speculated that the "shifted" or displaced nature of the PNR shown in Fig. 9 provides a barrier to the formation of a long-range ordered ferroelectric state in PMN at $T_c$.

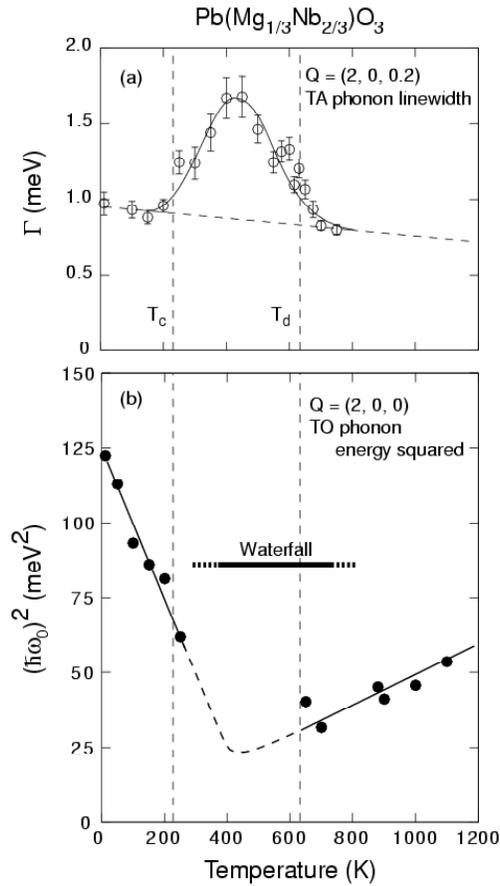

Fig. 10. Temperature dependence of the soft mode energy squared (lower panel) and the linewidth Γ of the TA mode (top panel). From Wakimoto *et al*. [12].

MODE COUPLING MODEL

Wakimoto *et al*. have very recently performed a quantitative analysis of the phonon lineshapes measured using neutron scattering techniques above the Burns temperature at 690 K in PMN [35]. Their findings indicate that the differences in the profiles and apparent energies of both the TA and TO phonons measured in the (200) and (300) Brillouin zones are well described by a simple model that couples the TA and soft TO modes. The primary parameter in this model is the q and T-dependent TO phonon linewidth $\Gamma(q,T)$. An example of how the TA phonon energy can appear to vary in different zones is shown in Fig. 11. The inset shows a systematic difference in the TA energies measured at (2,0,q) and (3,0,q), while the top and bottom panels highlight large differences in the intensities of both the TA and TO modes. The solid lines represent fits to the mode-coupling scattering cross section convolved with the instrumental resolution function. The data are clearly well described by the model in both the (200) and (300) zones.

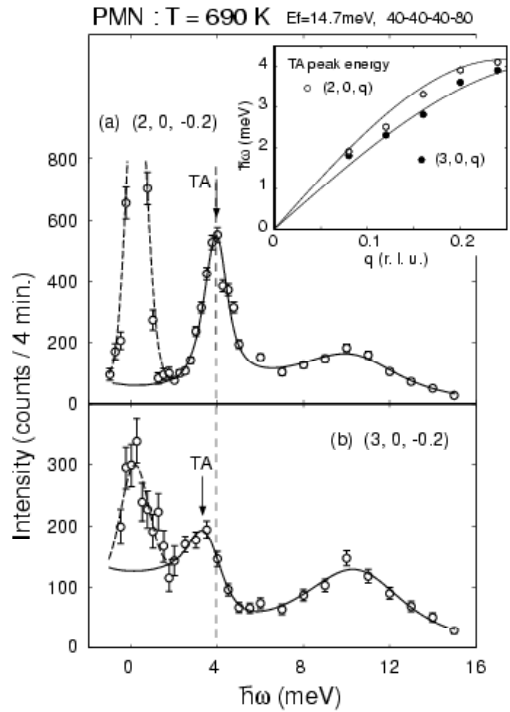

Fig. 11. Comparison of the TO and TA phonon profiles measured at (2,-0.2,0) and (3,-0.2,0). Notice the apparent shift of the TA phonon energy indicated by the vertical dashed line. From Wakimoto *et al*. [35].

This use of mode-coupling to describe phonon lineshapes in perovskite systems is not new. Harada *et al.* used a mode-coupling model to describe the asymmetric phonon lineshapes in BaTiO$_3$ in 1971 [36], while Naberezhnov recognized the importance of mode-coupling in PMN in their seminal paper in 1999 [14]. However in the context of Hirota *et al.*'s model of the displaced PNR, the mode-coupling approach brings a beautiful and unifying element to the ever-evolving picture of relaxor lattice dynamics. Namely, given that the soft TO mode is coupled to the TA mode, then it contains an acoustic component that, in turn, provides a natural origin for the uniform phase shift δ of the PNR when the soft mode condenses at the Burns temperature $T_d$. This idea is due to Y. Yamada [37].

FUTURE TOPICS

Two important topics on the basic research of relaxors have yet to be fully explored. The first one concerns a more definitive characterization of the polar nanoregions. The relationship between the atomic shifts below $T_d$ with the ionic vibrations of the soft TO mode was resolved by a proper analysis of the diffuse scattering [15]. However, we still have no unique picture of how these shifts are created. The mode-coupling approach of Wakimoto *et al.* [35] seems promising, yet it lacks the final connecting piece. Yamada and Takakura have proposed an alternative model that assumes the softening of the T1 acoustic mode along the [110] direction [38]. However experimental support for this assumption is lacking.

Another related question concerns the ratio of the volume of the PNR to that of the unshifted ferroelectric regions below $T_c$. There is a huge diffuse scattering below $T_c$, as pointed out long ago by Vakhrushev *et al.* [39]. In principle, the integrated diffuse scattering cross section expressed in units of the total scattering cross section of the crystal may give an estimate of this ratio. We would also like to know the change in size of the PNR between $T_d$ and $T_c$. This question is complicated by the fact that the diffuse scattering intensity depends not only on the volume of the PNR, but also on the magnitude of the PNR displacements.

Closely related to the questions about the PNR is the discovery of the new phase X in PZN by the neutron measurements of Ohwada *et al.* [31]. Very recently, a high-energy x-ray study by Xu *et al.* confirmed the presence of phase X in PZN after zero-field cooling [40]. The skin of this crystal does indeed transform into the expected rhombohedral phase. But the bulk of the crystal, a few microns below the surface, transforms into a new phase that is nearly cubic and only slightly tetragonally distorted. It is speculated that this phase X is caused by the persistence of the PNR to temperatures below $T_c$, and which therefore coexist with the regular ferroelectric state. Somehow, the PNR are unable to deform coherently into the rhombohedral phase without the help of an applied field oriented along the [111] direction. In this respect, it is interesting to note that PZN-8%PT between 510 K and 380 K manages to form a tetragonal distortion on cooling in zero field [31]. Further study of the diffuse scattering under different field-cooled conditions may provide the key to this problem.

ACKNOWLEDGMENTS

Work at Brookhaven National Laboratory is supported by the U.S. Department of Energy under Contract No. DE-AC02-98CH10886, and in part by the U.S.-Japan Cooperative Research Program on Neutron Scattering between the U.S. Department of Energy and the Japanese MONBU-KAGAKUSHO. We also acknowledge the support of the NIST Center for Neutron Research, the U.S. Department of Commerce, for providing the neutron facilities used in this work. We are deeply grateful to our many collaborators including D.E. Cox, L.E. Cross, K. Hirota, B. Noheda, R. Guo, K. Ohwada, S.-E. Park, P.W. Rehrig, C. Stock, Y. Uesu, S. Wakimoto, G. Xu, Y. Yamada, and Z.-G. Ye.